\documentclass[fleqn,10pt]{wlscirep}
\usepackage[justification=justified]{caption}
\usepackage[T1]{fontenc}
\usepackage{multibib}
\usepackage{svg}
\newcites{method}{Methods References}
\usepackage{amsmath}

\title{Quantum fluctuation of ferroelectric order in polar metals}

\author[1]{Fangyuan Gu}
\author[2]{Jie Wang}
\author[1,3]{Zi-Jian Lang}
\author[1,3,4,*]{Wei Ku}
\affil[1]{Tsung-Dao Lee Institute, Shanghai Jiao Tong University, Pudong, Shanghai 201210, China}
\affil[2]{Zhiyuan College, Shanghai Jiao Tong University, Shanghai 200240, China}
\affil[3]{School of Physics and Astronomy, Shanghai Jiao Tong University, Shanghai 200240, China}
\affil[4]{Key Laboratory of Artificial Structures and Quantum Control (Ministry of Education), Shanghai 200240, China}
\affil[5]{Shanghai Branch, Hefei National Laboratory, Shanghai 201315, China}
\affil[*]{weiku@sjtu.edu.cn}

\newcommand{\conc}{\elec\,/\,\text{f.u.}}

\newcommand{\elec}{e^-}

\DeclareMathOperator{\Tr}{Tr}

\begin{abstract}
Since its discovery a decade ago, ``polar metallic phase'' has ignited significant research interest, as it further functionalizes the switchable electric polarization of materials with additional transport capability, granting them great potential in next-generation electronic devices.
The polar metallic phase is an unusual metallic phase of matter containing long-range ferroelectric (FE) order in the electronic and atomic structure.
Distinct from the typical FE insulating phase, this phase spontaneously breaks the inversion symmetry but without global polarization.
Unexpectedly, the FE order is found to be dramatically suppressed by carriers and destroyed at moderate $\sim10\%$ carrier density.
Here, we propose a general mechanism based on carrier-induced quantum fluctuations to explain this puzzling phenomenon.
Basically, the quantum kinetic effect would drive the formation of polaronic quasi-particles made of the carriers and their surrounding dipoles.
The disruption in dipolar directions can therefore weaken or even destroy the FE order.
We demonstrate such polaron formation and the associated FE suppression via a simple model using exact diagonalization, perturbation, and quantum Monte Carlo approaches.
This quantum mechanism also provides an intuitive picture for many puzzling experimental findings, thereby facilitating new designs of multifunctional FE electronic devices augmented with quantum effects. 
\end{abstract}
\begin{document}
\flushbottom
\maketitle
\thispagestyle{empty}
%\linenumbers

Ferroelectric (FE) order corresponds to an ordering of local electric dipole moment in materials associated with a spontaneously broken inversion symmetry in the absence of an external electric field. %($\mathbf{E}_{ext}$)
Accordingly, insulating FE materials typically exhibit a field-switchable global spontaneous polarization ($\mathbf{P}$) and consequently a rather strong dielectric response.
This feature makes FE materials highly functional in electronic devices and other practical applications, including energy storage~\cite{Li2008,shen2015,lin_AppMatInt_2019}, photovoltaics~\cite{fahy_1994,2014Zenkevich,2015Sharma,wu_JACS_2019}, data storage and switching~\cite{arimoto_ishiwara_2004}.
In recent years, the attempt to functionalize FE materials with additional metallicity has stimulated intensive studies of the so-called ``polar metal'' phase in charge carrier-doped FE materials, which hosts metallic carriers in the presence of FE order.
%Yang2019, photovoltaics: Dharmadhikari1982, ,fridkin_ferroelectrics_2015 
%LiOsO3 Xiang_PRB2014,

Such a FE metallic ``polar metal'' state was first predicted by Anderson and Blount~\cite{anderson_blount}, who theorized broken inversion symmetry along a polar axis and the persistence of FE-like phase transitions in this metallic phase.
It was not until 2013 the polar metallic state was finally discovered in LiOsO$_3$ by Shi {\em et al.}~\cite{shi_nmat2013}.
Since then, many polar metals have been found in various carrier-doped FE materials, including perovskite oxides: BaTiO$_3$~\cite{kolodiazhnyi_PRL_2010,fujioka_2015,cordero_metal_2019, Zhou_commphys2019}, PbTiO$_3$~\cite{HXu_PRB2016, JGu_PRB2017}, CaTiO$_3$~\cite{Benedek_JMC2016}; NdNiO$_3$~\cite{Kim_nature2016}, LiOsO$_3$~\cite{shi_nmat2013, Laurita_natcomm2019}, Ca$_3$Ru$_2$O$_7$~\cite{Lei_nanolett2018}, and Cd$_2$Re$_2$O$_7$~\cite{Sergienko_prl2004}; hexagonal FE materials: LiGaGe~\cite{zhang_PRB2019} and LaAuGe~\cite{Du_APLmat2019}; and 2D layered materials: WTe$_2$~\cite{Fei_nature2018, Sharmaeaax5080} and MoTe$_2$~\cite{Sakaie_sciadv2016}.
% WTe2 calculation: ,yang_Phychemlett2018

\begin{figure}[!h]
\includegraphics[width=0.8\columnwidth]{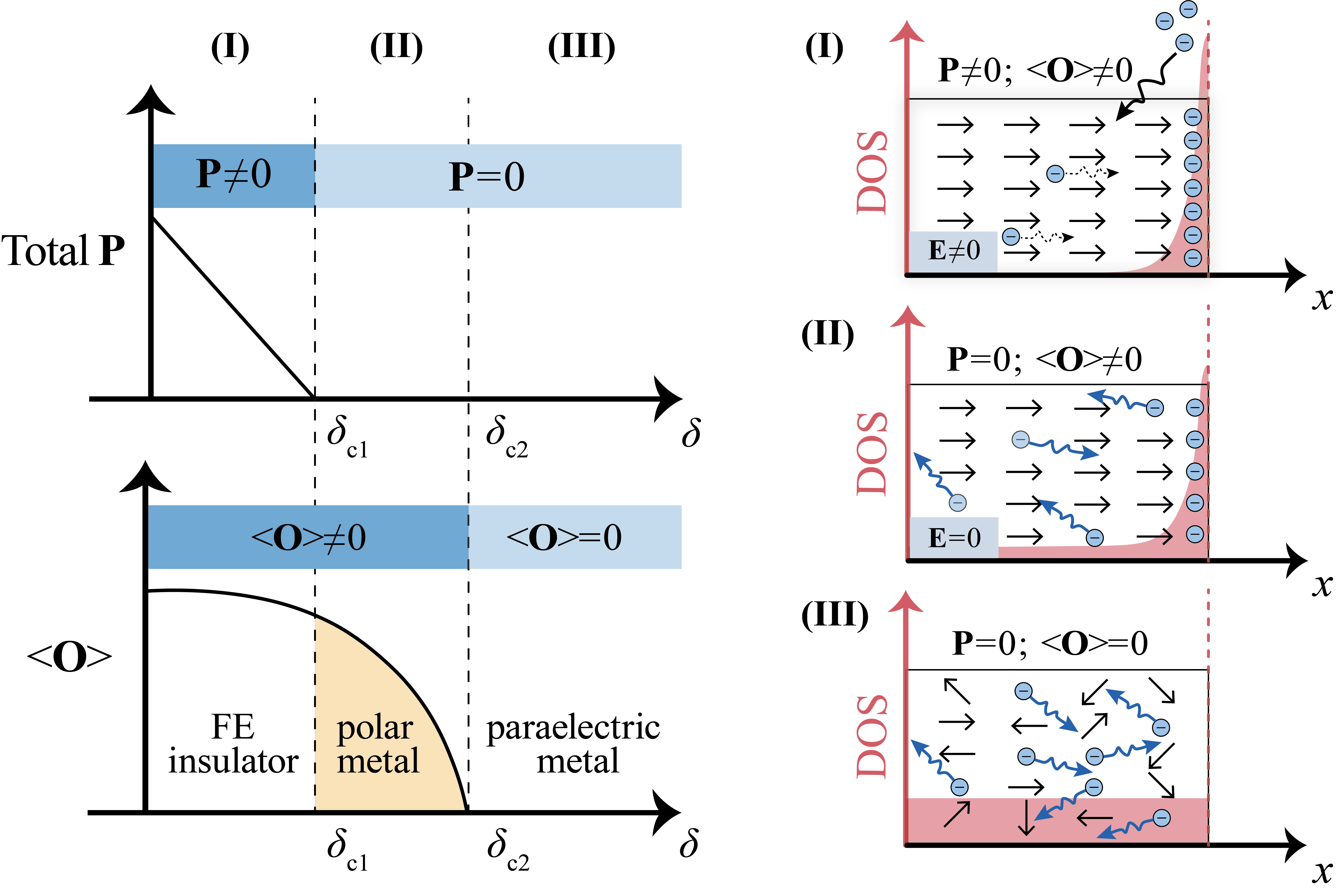}
%\includesvg[inkscapelatex=false,width=0.8\columnwidth]{Figures/phase_diagram.svg}
\centering
\caption{Left: Phase diagram of electron-doped FE materials. Right: Corresponding spatial distribution of electronic density of states and electric dipole at different stages of the phase diagram. 
%ranges of carrier density ($\delta$)
Stage I (FE insulator): With a tiny amount of doped electrons, charges are exhausted to screen the electric field $\mathbf{E}$ and the global dipole $\mathbf{P}$ by accumulating at the surface of the materials (or domain boundary).
Stage II (polar metal): With enough electrons to fully screen the electric field $\mathbf{E}$, some portion of the doped carriers can propagate in the system, while the FE order $\langle \mathrm{\mathbf{O}}\rangle$ corresponding to the non-centrosymmetric electronic and atomic structure remains.
%favors the highly-symmetric paraelectric (PE) phase,
%With intrinsic electric field shielded, every electron doped into the system affects a finite number ($\sim10$) of local dipoles;
%within a certain radius, and forms a quasiparticle, polaron; 
%the polarons merged and grow to the size of the whole system.
Stage III (paraelectric metal): With even more carriers introduced to the system, the FE order $\langle \mathrm{\mathbf{O}}\rangle$ is also destroyed and the surface charge is released to join the itinerant carriers.}
\label{fig:phase_diagram}
\end{figure}

In polar metallic phase, as shown in phase II of Fig.~\ref{fig:phase_diagram}, charge carriers can propagate freely in materials as soon as the global $\mathbf{P}$ and correspondingly the total electric field $\mathbf{E}$ is fully screened by $\delta_\textrm{c1}\sim 2\%$ of carriers accumulating on the domain boundaries and surfaces~\cite{Jirapa_FE_2009, Ivanchik_FE_1993}.
On the other hand, the FE order having a spontaneously broken symmetry actually still exists up to $\delta_\textrm{c2}\sim 10\%$~\cite{kolodiazhnyi_PRL_2010,fujioka_2015,cordero_metal_2019, HXu_PRB2016, Benedek_JMC2016, Lei_nanolett2018, Sergienko_prl2004}, despite the absence of the global $\mathbf{P}$.
Naturally, with the screening of the beneficial $\mathbf{E}$, ferroelectricity is expected to be weakened as widely found in current observations, for example, a decrease in phase transition temperatures ($T_c$) and coercive field ($E_c$)~\cite{HARDTL_1972,hwang_2010,kolodiazhnyi_PRL_2010,fujioka_2015,cordero_metal_2019,kolodiazhnyi_PRB_2008}, a remarkably reduced off-center FE distortions~\cite{fujioka_2015,cordero_metal_2019,kolodiazhnyi_PRL_2010}, a softening of the soft mode phonon~\cite{hwang_2010} and an emergence of the over-damped highly-anharmonic central mode~\cite{hwang_2010,bellaiche_2008}.
%,Petzelt2008

There are however, many unexpected puzzling behaviors in this phase, associated with the introduction of metallic carriers, including an anomalous sign reversal in the Hall coefficient in $n$-doped $\mathrm{Ba}\mathrm{Ti}\mathrm{O}_{3}$ single crystal~\cite{kolodiazhnyi_PRB_2008}, a remarkably low carrier scattering rate, a modest intrinsic carrier mobility~\cite{zhu2015}, and a sudden increase in the real part of the dielectric function in \textit{sub}-THz region in lead halide perovskites~\cite{Wang_2021jacs, Kiyoshi_2018_NatMat}.
Even more unusual is that the observed transition temperature of the lowest-temperature FE phase appears to be nearly doping independent~\cite{kolodiazhnyi_PRL_2010}, or even slightly increasing with doping in $n$-doped $\mathrm{Ba}\mathrm{Ti}\mathrm{O}_{3}$~\cite{cordero_metal_2019}, despite the overall weakening of the FE order $\langle \mathrm{\mathbf{O}}\rangle$.

Still, the most puzzling is why in this phase the FE order can be so efficiently suppressed by merely $\sim10\%$ of doping, particularly when the global $\mathbf{P}$ and $\mathbf{E}$ are already fully compensated in the entire phase.
%density ($x_\textrm{c}\sim 0.023-0.125\;\conc$) 
%However, puzzles remain in the intermediate stage, which is the polar metallic state. This is the stage where the puzzle remains.
As shown in Fig.~\ref{fig:phase_diagram}, apart from accumulating at the FE surface/domain boundary to screen the $\mathbf{E}$ field, the carriers also propagate in the field-free region in this phase.
Intuitively, when residing in a small atom, each carrier can enlarge the atomic size and thus remove the local polar distortion.
However, it is not obvious how merely $\sim10\%$ of depletion of local dipoles can destroy so effectively the long-range order $\langle \mathrm{\mathbf{O}}\rangle$ of the entire system.
This strange phenomenon clearly reflects the fundamental nature of the polar metallic state.
A proper microscopic understanding of it would surely provide the basis for a natural explanation of other puzzles above and pave the way for further engineering and optimization of these functional materials.

This puzzling behavior, however, poses a clear challenge to current pictures of polar metallic phase.
The common mesoscopic classical picture of nanometer-scale domain mixture~\cite{fujioka_2015} and the observation of diffusive phase transition~\cite{chakraborty_ray_2014,jinhuo_2016} provide no mechanism directly, particularly considering the rather small amount of impurities and the associated disorder effects.
%, Liu2018
Another popular scenario that leads to successful geometric design~\cite{Kim_nature2016, Benedek_JMC2016, Filippetti_natcomm2016}, the so-called ``weak coupling hypothesis''~\cite{Puggioni2014_natcomm} between the carrier and the FE order, is obviously not applicable to address the efficient destruction of the latter via the introduction of the former.
Similarly, density functional studies~\cite{Spaldin_JchemC2021} assume very large time-scale separation of electron and lattice dynamics, directly contradicting the observed really low carrier mobility~\cite{zhu2015}.
Even more specific picture aiming to address this particular issue, for example, consideration based on carriers' screening of supposedly beneficial long-range interaction~\cite{Benedek_JMC2016} encounters difficulty since the long-range interaction was shown unnecessary to establish a stable FE order~\cite{Senn_PRL_2016, wang_PRL_2012}.
In fact, classical pictures would generically have fundamental difficulty circumventing the thermodynamical requirement of entropy reduction at low temperatures, which instead promotes ordering even with the introduction of itinerant carriers.

\begin{figure}[t!]
\includegraphics[width=0.75\columnwidth]{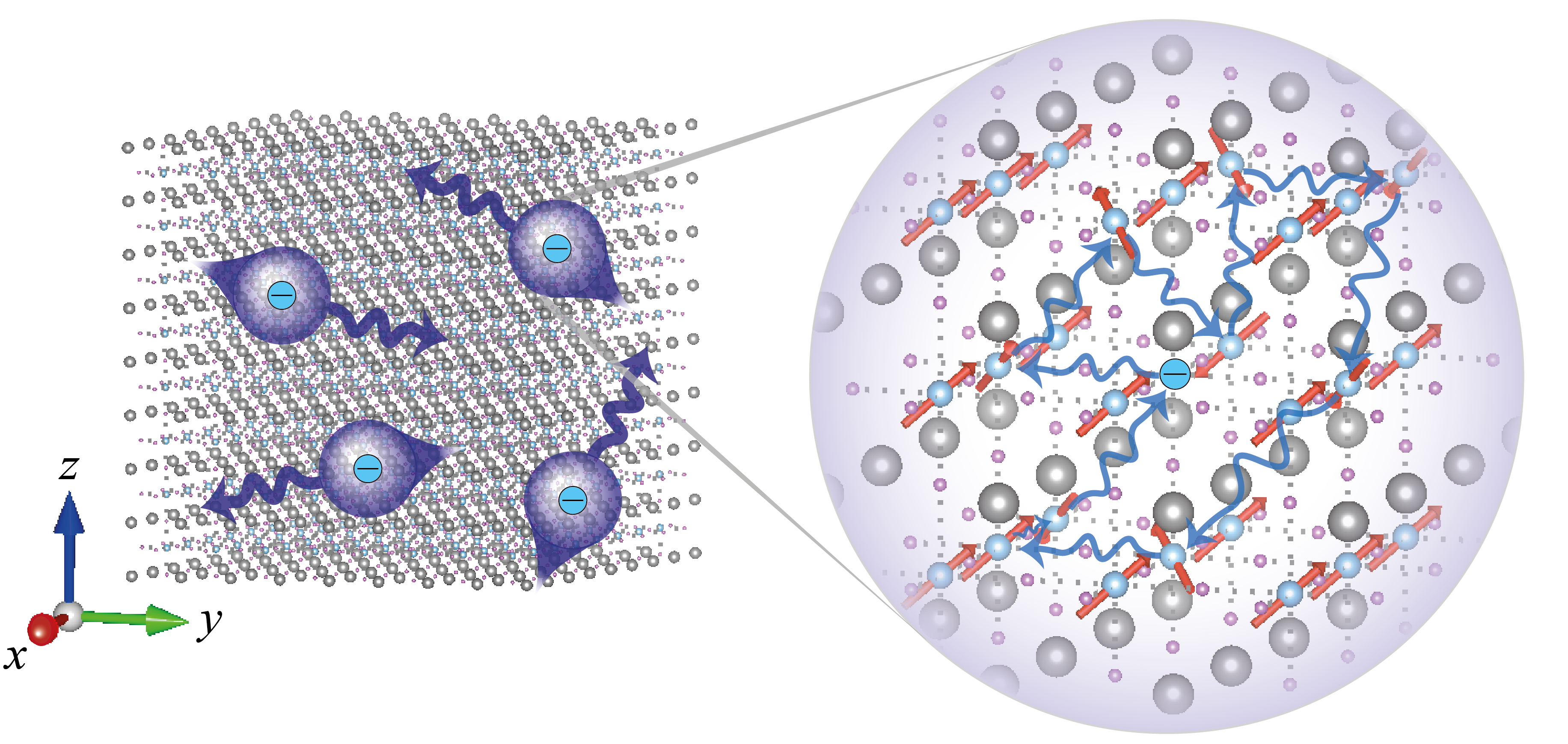}
%\includesvg[inkscapelatex=false,width=0.75\columnwidth]{Figures/polaron_demo.svg}
\centering
\caption{
Schematic of polarons. In polar metallic phase, itinerant low-energy carriers are slow emergent quasi-particles named ``polarons'' made of charge centers and fluctuating local dipoles surrounding them.
Such polarons acquire their internal structure via fast dynamics of the charge that disrupts the orientation of nearby local dipoles, a process commonly referred to as quantum fluctuation.
%In quantum mechanics, this is represented via superposition of various charge/dipole configurations.
%messed up
%With the quantum superposition of these local dipole states, the FE order would be effectively suppressed.
%As a dressed quasiparticle, polarons move with much slower dynamics compared to bare carriers and leave trails behind them.
}
\label{fig:polaron_formation}
\end{figure}

%(formation and property of polaron)
Here, we propose a general mechanism for the efficient suppression of the FE order in the polar metallic phase through its \textit{quantum} fluctuation, associated with the generic formation of itinerant ``polarons''.
As illustrated in Fig.~\ref{fig:polaron_formation} and quantified below, the kinetic process of quantum carrier that allows it to move between atoms can naturally lead to a superposition of disoriented dipoles in its vicinity.
Such a high-energy (fast) process would ensure a rigid local structure of the carrier and its surrounding disturbed dipoles at low energy (longer time scale) relevant to the transport properties or broken symmetry phase of these polar metallic materials.
It is therefore convenient to regard the resulting local structure as a new emergent mobile particle named polaron.
Since the dipoles are disrupted within, large mobile polarons can thus efficiently weaken or destroy the long-range FE order even at low carrier density, as observed experimentally.
In great contrast to the classical pictures, the quantum superposition of disoriented dipoles indicates multiple possible directions of each dipole even at zero temperature, since a coherent superposition carries no internal entropy.
We demonstrate below the formation of such a quantum polaron and its disrupting effects on FE order with a simple model using exact diagonalization, perturbation, and world-line quantum Monte Carlo.
The proposed quantum mechanism can offer natural explanations to many other anomalous experimental findings in polar metals.
Particularly, the explicit inclusion of quantum physics should prove essential in understanding and engineering polar metals in general.

\vspace{10pt}

\begin{figure}[t]
\includegraphics[width=0.93\columnwidth]{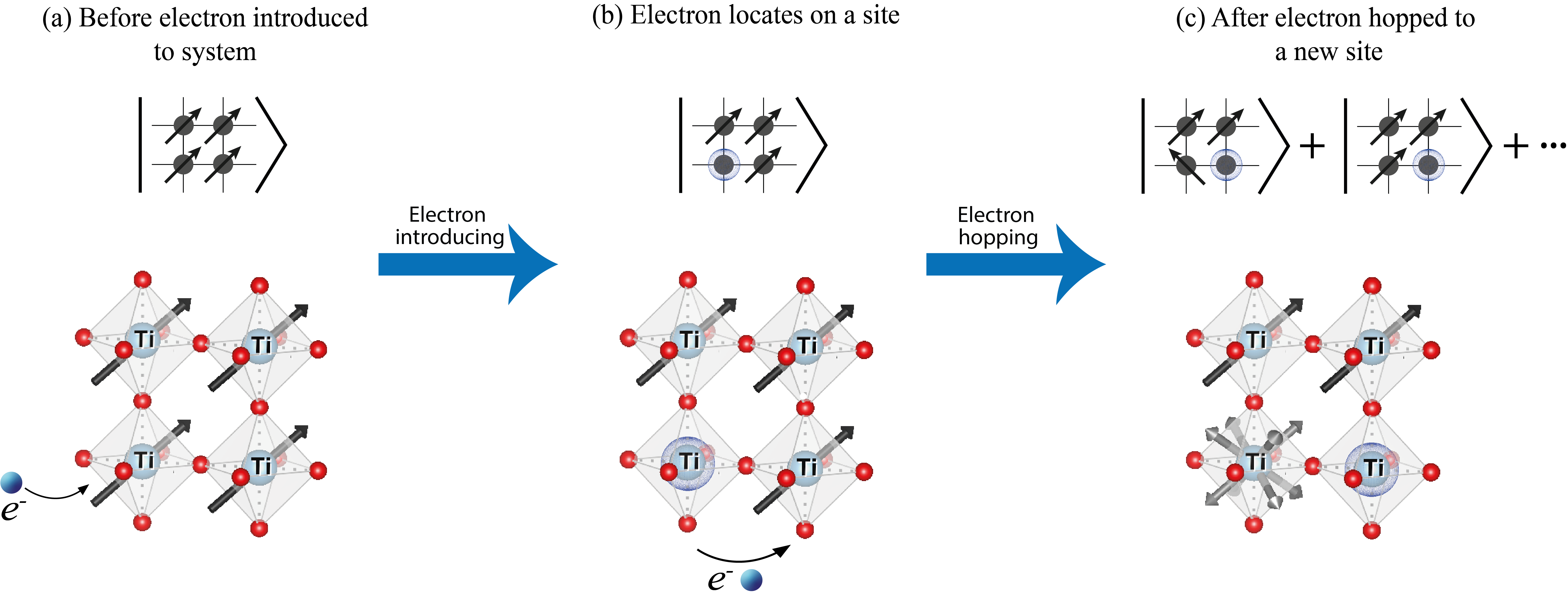}
%\includesvg[inkscapelatex=false,width=0.93\columnwidth]{Figures/hopping_schematic.svg}
\centering
\caption{Illustration of the quantum kinetic process in our model. (a) Before the electron is introduced, the local dipoles are well aligned along the ordered direction. (b) Upon introduction of an electron, the local octahedron containing one more charge will be in the charged mode without a local dipole. (c) When the electron propagates to the neighboring octahedron, it leaves behind a dipolar mode in one of the possible directions. As shown in the upper panel, a quantum polaron state contains linear superposition (illustrated by multi-directional arrows in the lower panel) of all possible such quantum fluctuations in its vicinity. Note that in our model the local state of an octahedron is simplified to either a charged mode or one of the eight dipolar modes with different directions.
%of (a) a single electron introducing to the FE system, (b) the hopping process of the itinerant electron, and (c) the effects of local dipoles when the electron arrives at a new site.
%The interactions between a single itinerant CBE and the local dipoles are shown in the diagram.
}
\label{fig:hopping_schematic}
\end{figure}

%\nolinenumbers
\section*{Theory and Model}\label{sec:theory_model}
%\linenumbers
To demonstrate the quantum mechanical formation of polaron in carrier-doped FE materials and its essential properties, let us consider a general strong coupling model as illustrated in Fig.~\ref{fig:hopping_schematic}.
In the absence of itinerant carrier as in Fig.~\ref{fig:hopping_schematic}(a), the local atomic and electronic structure at a particular atomic site $i$ are assumed to have deformed away from a fully symmetric one due to some high-energy physical mechanism and thereby host a local electric dipole mode $n$ in one of the energetically favored directions with a fixed dipole size $p_0$, for example, $n=1,2,3\cdots 8$ corresponding to one of the eight $\langle 111\rangle$ directions, $\mathbf{\hat{n}}$.
We label such a local dipole mode via a second-quantized creation operator $a_{in}^{\dagger}$.
Upon introduction of an itinerant carrier (denoted by $c_{i}^{\dagger}$) to such a local unit as in Fig.~\ref{fig:hopping_schematic}(b), the additional charge would greatly enlarge the size of the central cations and consequently remove such local deformation and thus the local dipole mode as well.
The simplest but rather generic effective Hamiltonian then reads,
%\begin{linenomath*}
\begin{equation}
\begin{aligned}
    H &=H_t+H_{MD}+H_{DD}+H_R \\
    &=\sum_{\langle ii'\rangle nn'}t_{ii'}c_i^{\dagger}c_{i'}a_{i'n'}^{\dagger}a_{in}+\sum_{ii'n'}{K}_{i,i'n'}c_i^{\dagger} c_i a_{i' n'}^{\dagger}a_{i'n'}-\sum_{ii'nn'}{J}_{in,i^\prime n^\prime}a_{in}^{\dagger} a_{in} a_{i' n'}^{\dagger}a_{i'n'} +\sum_{inn'}{R}_{nn'}a_{in}^{\dagger} a_{i n^\prime},
\end{aligned}
\label{eqn:total_hamil}
\end{equation}
%\end{linenomath*}
where $c_i^{\dagger}(c_i)$ and $a_{in}^{\dagger}(a_{in})$ follow fermionic and bosonic statistics, respectively.
Since each site can only host either one of the dipole modes or one with a carrier, they follow the strict single-choice exclusion condition,
%\begin{linenomath*}
\begin{equation}
    \begin{cases}
     \   c_i^{\dagger}a_{in}^{\dagger}=a_{in}^{\dagger}a_{in'}^{\dagger}=0\\[10pt]
    \   c_i^{\dagger}c_i + \sum_{n}a_{i n}^{\dagger}a_{i n}=1.
    \end{cases}       
\end{equation}
%\end{linenomath*}

The third term $H_{DD}$ of Eq.~\ref{eqn:total_hamil} is simply the second quantized representation of the familiar inter-site dipole-dipole coupling, $H_{DD}=-\sum_{i,i'}\left(J_{ii'}/p_0^{2}\right)\mathbf{p}_i\cdot\mathbf{p}_{i'}$, between dipoles $\mathbf{p}_i=p_0\sum_n \mathbf{\hat{n}} a_{in}^{\dagger} a_{in}$.
Similarly, the second term $H_{MD}$ represents the monopole-dipole coupling, $H_{MD}=-\sum_{i,i'}\left(K_{ii'}/p_0\right)\mathbf{\hat{r}}_{ii'}\cdot\mathbf{p}_{i'}$, between the charged carrier and the surrounding dipoles~\cite{Kiyoshi_2018_NatMat} along the direction of relative position $\hat{\mathbf{r}}_{ii'}=\left(\mathbf{r}_i-\mathbf{r}_{i'}\right)/|\mathbf{r}_i-\mathbf{r}_{i'}|$.
The last term $H_{R}$ describes the intrinsic fluctuation between dipole modes corresponding to the switch of the local dipole directions.

The novel physics introduced by itinerant carriers is mostly through its quantum mechanical kinetic effect given by the first term, $H_t$.
It describes the hopping of an itinerant carrier from a site $i'$ to a neighboring site $i$ and thereby removing the dipole at site $i$.
This process also leaves behind a site $i'$ without carrier, which would then develop a dipole.
A key aspect of quantum mechanics is that the newly developed dipole can in principle be in any possible modes $n'$, or more generally in a quantum superposition of them shown in Fig.~\ref{fig:hopping_schematic}(c).
As to be demonstrated in the results below, this feature is the essential ingredient for the quantum fluctuation of FE order inside the polarons.

Note that Eq.~\ref{eqn:total_hamil} is specifically meant to capture the eV-scale physics that establishes robust local polarons.
It therefore does not include many of the low-energy ($sub$-eV scale, slow) processes, for example, the small fluctuation of dipoles near each stable mode due to the 20 meV-scale phononic vibration of the atoms~\cite{Kozina_2019Natphys}, or even slower $\sim$4 meV-scale domain wall dynamics of depolarization field~\cite{Zhao_2019Natcomm}.
For a similar reason, very high-energy (multiple-eV scale, rapid) dynamics beyond the scale of polaronic formation have been conceptually decoupled from Eq.~\ref{eqn:total_hamil} through renormalizing the remaining physical effects in Eq.~\ref{eqn:total_hamil}.
For example, one might wonder about the dynamical process corresponding to the disappearance/emergence of local dipoles associated with charge addition/removal to the local transitional metal sites.
However, since this involves a large beyond-eV scale Coulomb energy change, for the eV-scale physics described by Eq.~\ref{eqn:total_hamil} the influence of very high-energy physics is absorbed and contributes to the strength of the remaining \textit{effective} parameters $t$, $K$, $J$, and $R$.

%(brief methods, results)
%Below we will study the structure of a single polaron using three numerical calculations: exact diagonalization, perturbation theory, and world-line quantum Monte Carlo (QMC).
%The first two provide a clear picture of the polaron formation and kinetic energy dependence of the polaron size, while the third reveals additional effects of temperature and polaron motion on quantum fluctuation.
%The results demonstrate consistently that the local dipoles inside the polaron host quantum superposition are along various directions and thus not well aligned with the FE order.
%Furthermore, the polaron size is found to grow rapidly as the kinetic energy increases toward a critical value.
%Together, such a quantum fluctuation can therefore suppress efficiently the FE order with sufficiently strong kinetic energy, even with just $\sim10\%$ small carrier density.

\begin{figure}[!t]
  \centering
  \includegraphics[width=\columnwidth]{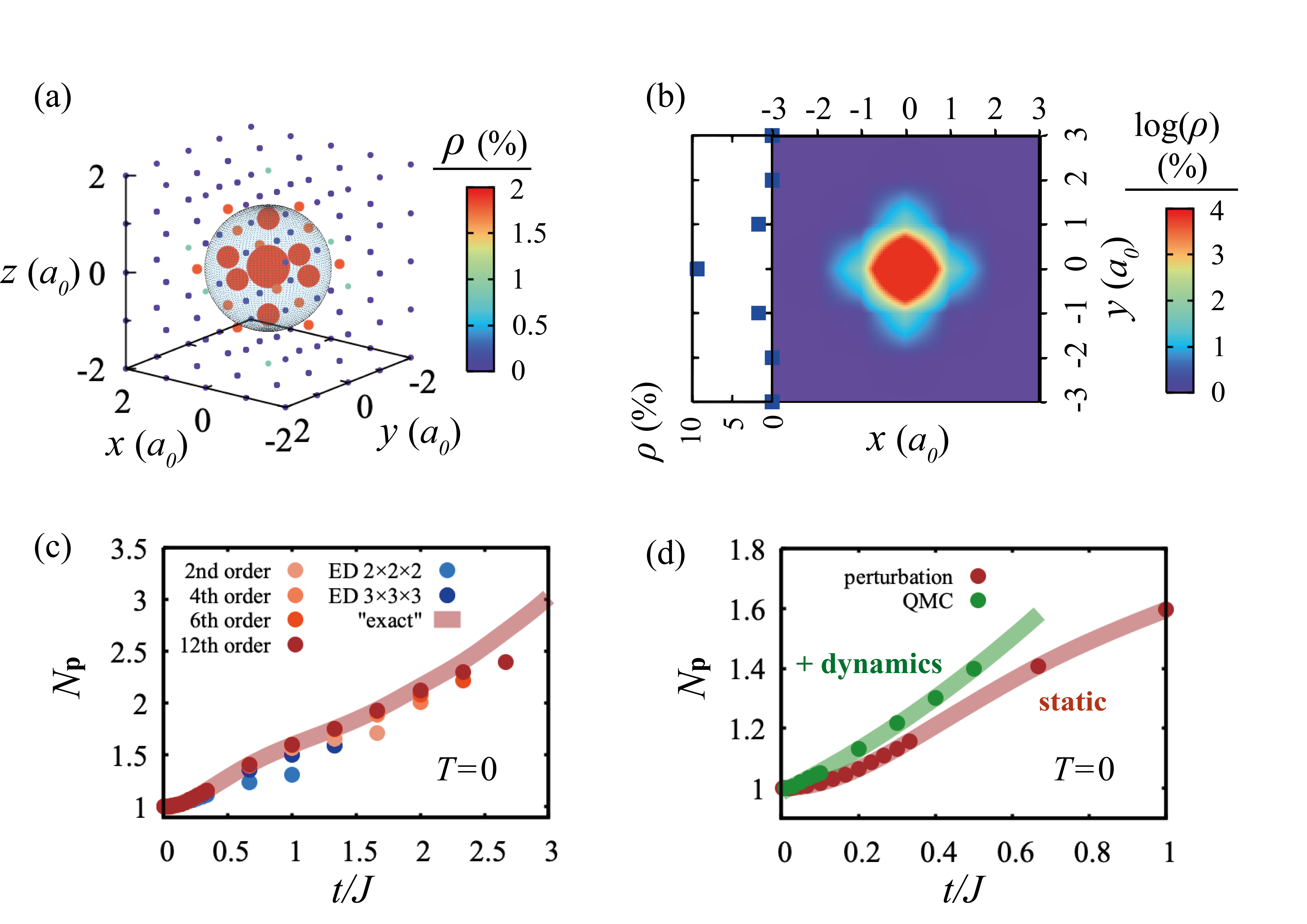}
% \includesvg[inkscapelatex=false,width=\columnwidth]{Figures/fig4.svg}
\caption{
Structure of the polaron and its disruption on the local dipoles.
(a) The site-distribution of charge density, $\rho_i=|\langle i|\psi \rangle|^2$, of a polaron with $t/J=2$ in unit of percentage.
(b) The same along the (0,-3,0)-to-(0,3,0) path and in the $(x,y,0)$ plane.
Note that a log scale is employed for the latter for better visualization.
(c) Kinetic strength, $t/J$, dependence of the average suppressed number of dipoles, $N_{\mathbf{p}}$, calculated using ED and perturbation approaches up to the 12th order.
The thick ``exact'' line is estimated as a guide to the eye.
(d) Enhancement of $N_{\mathbf{p}}$ due to additional slower itinerant dynamics of the polarons, illustrated by the $t/J$ dependence of $N_{\mathbf{p}}$ of a dynamic polaron (through the world-line QMC) with respect to that of a static one (via the perturbation approach).
Results from the finite-temperature world-line QMC calculation are extrapolated to the $T=0$ quantum limit.
}
\label{fig:dipole_result}
\end{figure}

%\nolinenumbers
\section*{Formation, internal structure, and effect of quantum polarons} \label{sec:results_discussion}
%\linenumbers
Fig.~\ref{fig:dipole_result}(a) demonstrates the structure of a quantum polaron from the ground state of our exact diagonalization and perturbation calculation.
It shows a density distribution of a quantum state describing the real charge carriers relevant to the slow transport process.
This carrier has a well-defined location, and its density extends to surrounding atoms.
Importantly, the local electric dipoles within the scope of this extension are disrupted from the FE order $\langle \mathrm{\mathbf{O}}\rangle$ (so-called ``quantum fluctuation'') due to the superposition illustrated in Fig.~\ref{fig:hopping_schematic}(c).
In other words, polarons indeed would form in a polarizable media consisting of particles and their nearby disrupted local dipoles.

Physically, this particular superpositioned polaronic structure can be understood as follows.
Within the short time scale inverse proportional to $t_{ii^\prime}$ of the quantum kinetic effect $H_t$, electrons can rapidly hop between different sites.
Thus from the perspective of a longer time scale relevant to the transport properties, such fast motion would lead to various probabilities of leaving behind trails of disordered dipoles around an average center (c.f. right panel of Fig.~\ref{fig:polaron_formation}.)
Naturally, these rapid processes are impossible to decipher using slower probes and thus can be regarded as part of the rigid internal structure of a new quasi-particle named polaron.
The coherent quantum superposition is merely the mathematical representation to encode partially the dynamics of these rapid processes, including the density distribution.
%In short, a polaron will form consisting of a particle and disrupted local dipole in a polarizable media.
%Looking into the internal structure of the finite size polaron, 
Following the smooth decay of the carrier density away from the center, as shown in Fig.~\ref{fig:dipole_result}(b), the average internal structure of polaron would correspond to a smooth distribution of transitional metal-oxygen bond length, on top of the typical long-short bond length pair outside the polarons, in good agreement with the structural refinement via neutron scattering~\cite{jeong_2011}.

\begin{figure}[!t]
\includegraphics[width=0.55\columnwidth]{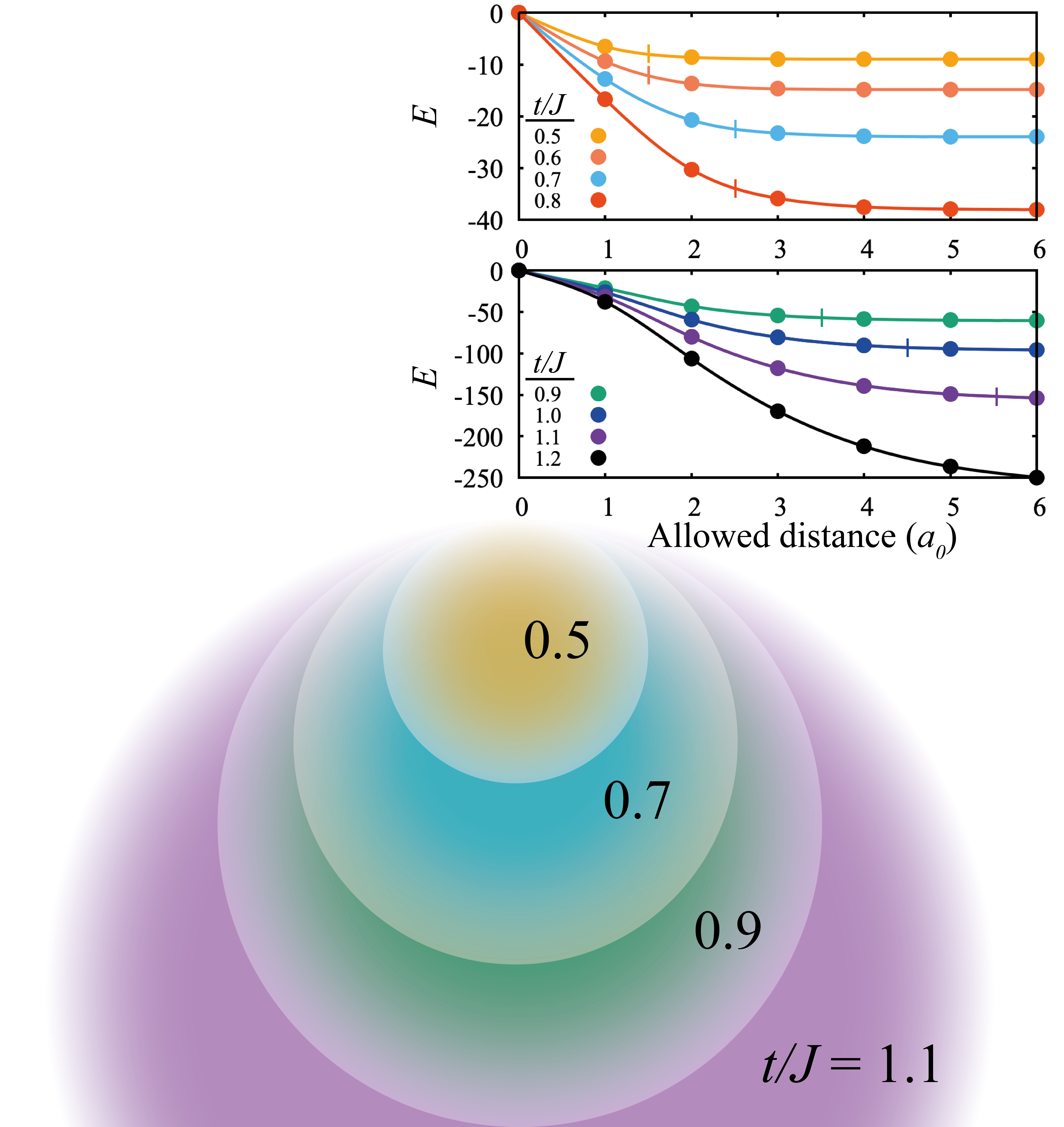}
%\includesvg[inkscapelatex=false,width=0.55\columnwidth]{Figures/polaron_size.svg}
\centering
\caption{
Superlinear growth of polaron radius with respect to the kinetic strength of the bare carrier, $t/J$.
The insets show the lowering in energy, $E(r)$, with the allowed distance $r$ for a polaron to grow given each $t/J$ parameter in our perturbation calculation.
(See the Methods section for the convergence analysis.)
From this, the radius of the polaron (denoted by the short vertical lines) in the presence of competing low-energy physics (of strength $4J$ for example) can be estimated as the bound at which the additional energy gain from growing one step larger $E(r+a_0) - E(r)$ is insufficient to overcome the competition.
%Note that the $\Delta E_{\mathrm{max}}$ is the maximum energy correction of the allowed distance of the carrier at a certain $t/J$. 
%According to the perturbation convergence, the energy correction of the $j$-th hopping step is roughly proportional to $t^{j}/\left((j)!(6J)^{j-1}\right)$, which rapidly converges with the distance with small kinetic strength;
%while with a larger $t/J$, the energy correction, $\Delta E$, increases to a maximum value at a certain distance and then converges gradually.
%The energy gain vanishes as the allowed distance is larger than the farthest site that the carr
%The additional energy gain ($\Delta E$) within the distance of the farthest octahedral site that the carrier could reach in its virtual dynamical process.
%This  indicates the rough size of the polaron.
%Larger kinetic strength $t/J$ naturally leads to larger polaron.
%Details of the convergence of the perturbation approach to the higher-order terms with different values of $t/J$ are given in the Methods section.
}
\label{fig:convergence_size}
\end{figure}

Since the quantum polaron forms as a result of the rapid kinetic processes associated with $t/J$, naturally the stronger the kinetic process is, the larger the polaron becomes.
Fig.~\ref{fig:convergence_size} shows that the energy, $E$, decreases as the kinetic process extends to a longer distance.
For a fixed $t/J$, initially a significant energy gain $|E|$ can be obtained by allowing the kinetic process to cover a larger region.
After a characteristic distance, the gain starts to diminish such that the corresponding coherence can be challenged by other low-energy physics.
One can therefore associate this characteristic distance as the radius of the polaron.
Fig.~\ref{fig:convergence_size} shows that not only a stronger $t/J$ would indeed lead to a larger quantum polaron, but also the effect is beyond linear.

As an essential characteristic, the size of the polaron directly affects its ability to locally disrupt the FE order $\langle \mathrm{\mathbf{O}}\rangle$.
Fig.~\ref{fig:dipole_result}(c) shows the reduction of average local dipoles around a single polaron in unit of number of local dipoles, $N_{\mathbf{p}}=\sum_i\left( 1-\left(\langle{\mathbf{p}_i\rangle}/p_0\right)\cdot \left(\mathbf{P}/|\mathbf{P}|\right)\right)$, obtained from our calculations.
In the weak kinetic limit, $t/J \rightarrow 0$, the polaron is tightly bound to a single atomic site and thus it removes only one dipole at that site $N_{\mathbf{p}} \rightarrow 1$.
As the kinetic strength of the bare carrier ($t/J$) grows, the size of the polaron increases and is thus able to damage more effectively the surrounding local dipoles, i.e. an increasing $N_{\mathbf{p}}$.
(Notice that as the polaron grows in size, it naturally requires a larger system size, or higher order of perturbation, for the calculation to reach a fully converged $N_{\mathbf{p}}$, as illustrated by the thick transparent lines.)

With such a rapid growth of disruption, the experimentally observed efficient destruction of FE order can now be intuitively understood.
Even with a relatively low density such as $\delta \sim 10\%$, as long as the polaron size grows to the percolation threshold $\sim 1/\delta$ under strong enough kinetic processes, the remaining local dipoles would be disconnected by the polarons and unable to align their directions.
In that case, given a strong dipolar formation energy~\cite{jeong_2011, Stern2004}, the system would enter a paraelectric phase containing disordered local dipoles.
From a general consideration of entropy, such a low-temperature paraelectric phase with disordered local dipoles is nearly impossible to realize classically, but rather natural given that quantum superposition of different dipolar directions carries no internal entropy.
%Zalar2003PRL,

In addition, the above disruption of FE order should be further enhanced by the slow itinerant dynamics of the polaron.
Fig.~\ref{fig:dipole_result}(d) shows that at larger $t/J$ our resulting $N_{\mathbf{p}}$ from the QMC calculation becomes systematically larger than that of a single immobile polaron obtained above.
This is because, in the QMC calculation the polaron can propagate in the system, thus introducing additional dynamical (time-dependent) quantum fluctuation of the electric dipoles.
Consequently, such a dynamical effect can further increase the efficiency in suppressing the long-range FE order and thereby lower the critical carrier density $\delta_{c2}$ in Fig.~\ref{fig:phase_diagram}.
(Note that such an itinerant dynamics is hard to circumvent, since in real materials disorder potentials are typically of $sub$-eV scale and therefore insufficient to induce real localized states~\cite{Ku_2012PRL}, despite the local screening density around the charged impurities~\cite{kolodiazhnyi_PRL_2010}.)
%as commonly proposed~\cite{choi_APL_2011, tsunoda_PRM_2019}

A less intuitive effect of such quantum fluctuation is the enhanced stability of the ground state despite a reduced order parameter.
This is clearly indicated by an enlarged energy splitting between the ground state and the excited states due to the quantum fluctuation in our calculation, as expected from the level-repulsion principle of quantum mechanics.
Consequently, compared with the undoped systems that contain no itinerant carrier, the transition temperature of the lowest-temperature phase can instead \textit{increase} slightly, even though its order parameter is smaller.
This is in great contrast to the typical effect of thermal fluctuation that associates a smaller order parameter with a lower transition temperature.
Interestingly, such a counter-intuitive effect has actually been observed in the elastic measurement~\cite{cordero_metal_2019} and the resistivity measurement~\cite{kolodiazhnyi_PRB_2008} of the lowest-temperature orthorhombic (Amm2) to rhombohedral (R3m) phase transition at 183K, showing a slight $\sim$10K increase upon $\delta=0.0354\;\conc$ doping.
%(\delta=6.1*10^19 resistivity)
Similarly, in many prototypical polar metallic materials, one finds a much weaker doping reduction of the transition temperature for lower-temperature transition~\cite{kolodiazhnyi_PRL_2010,fujioka_2015}.
Such an unusual trend is hard to explain via thermal fluctuation, but is natural from the enhanced stability associated with quantum fluctuation.

An important consequence of the polaron formation is a serious enhancement of the effective mass of the carriers and suppression of their mobility.
This is because these polaronic carriers are heavily dressed by quantum fluctuation involving not only the charge fluctuation but also the dynamics of the polarizable medium around its center.
For example, the effective mass can be enhanced by more than two orders of magnitude at $t/J < 0.1$, when the environment is strongly polarizable.
Such an enormous mass enhancement has in fact been observed in various experiments~\cite{zhu2015, kolodiazhnyi_PRB_2008}.
Note that our proposed mechanism is capable of slowing down the carrier dynamics from eV to 10 meV or even meV scale.
This is in great contrast to the currently proposed polaronic pictures employing slow~\cite{Kiyoshi_2018_NatMat,  PNAs_Wang2022, Bonn_2017ACS} (or even static~\cite{Wang_2021jacs, Nanolett_ma2015}) phonon modes, which are relevant only if the carrier dynamics are of a similar time scale to that of the phonons.
Our quantum fluctuation-induced polaron formation offers a high-energy mechanism to slow the carriers down significantly such that further dressing the polaron through these lower-energy polaronic mechanisms can become effective.
%Miyata_2017Sciadv,

%\nolinenumbers
\section*{Summary}\label{sec:summary}
%\linenumbers
To explain the puzzling effective suppression of FE order in polar metals through slight carrier doping, we propose a general mechanism of polaron formation based on the quantum fluctuation of carriers in a highly polarizable medium.
We first demonstrate the formation of polarons as the emerged slow carriers that absorbs the faster dynamics into their internal structure through quantum superposition of states with disordered electric dipoles nearby.
We then find that the size of polaron is controlled mainly by the underlying kinetic processes, such that a large polaron can easily form in reality to cause the observed efficient suppression of FE order.
This leads naturally to the low-temperature quantum paraelectric phase indescribable by classical physics.
Consistent with the observed heavy mass of the carriers, the remaining polaron dynamics can be orders of magnitude slower than the underlying kinetic processes of energy as high as eV-scale.
Finally, our QMC calculation indicates that the slow polaron dynamics further suppresses the FE order.
Our proposed mechanism provides the essential foundation for previously proposed slow polaronic mechanisms and sets the basic framework for a generic description of polar metals.

%\nolinenumbers
\bibliography{polaron_paper}
\vspace{9pt}

%\nolinenumbers
\section*{Methods}\label{sec:method}

%\linenumbers
We study the formation and the static/dynamic properties of the polaron induced by an itinerant carrier in the ground state long-range FE-ordered perovskite through three approaches: (i) Exact diagonalization (ED), (ii) Perturbation theory, and (iii) World-line quantum Monte Carlo (QMC).
In the analysis, we aim at demonstrating the generic features and physical trends in the strongly correlated regime, when the effects of kinetic energy are weaker than those of the near neighboring interaction.
Since in this strongly correlated regime the most essential physical trends are dominated by how quantum kinetic energy adapts to the constraint of strong local interactions $H_{DD}$ and $H_{MD}$, we simplify our discussion by reducing the interactions to only the nearest neighboring cooperative dipolar interaction~\citemethod{Bellaiche_2013PRB} ${J}_{in,i^\prime n^\prime}=J \ \mathbf{\hat{n}}\cdot\mathbf{\hat{n}'}$ with nearest neighboring $J=0.7$ eV~\citemethod{zhong_PRB_1995}, and explore the parameter $t_{ii'}=t$ for the nearest neighboring hopping ranging from $0^+$ to 6$J$.
While the introduction of $H_{MD}$~\cite{Kiyoshi_2018_NatMat} and long-range $H_{DD}$ can slightly modify the internal shape of the polarons, it would not affect the qualitative trend since their energy is much smaller than that of the cooperative ligand displacement~\citemethod{Bellaiche_2013PRB}.
Similarly, given the already strong quantum fluctuation due to the eV-scale kinetic energy, it is convenient to also drop the much weaker local fluctuation $H_{R}$, considering the small $\sim$20meV potential barriers between the local dipole modes in real materials~\citemethod{Gu2021}.
\begin{enumerate}[label=(\roman*)]
    \item \textit{Exact diagonalization} (ED): In this study, we use an ED calculation to provide accurate results for relatively small polarons at small $t/J$ values.
We start with a pure FE-ordered system with a single carrier introduced into the system as schematically illustrated in Fig.~\ref{fig:hopping_schematic}.
%Then, let us consider the hopping process of the doped carrier.
%which is discussed in \textit{Theory and Model} section.
Regarding the hopping of the doped carrier in a 3D bulk system, there are six equivalent directions of the nearest neighboring unit cells for the carrier to hop to.
On the site that the carrier left from, there are eight possible modes orients along $\langle 111\rangle$ of the local dipole moment.
Therefore, the dimension of the Hilbert space corresponds to the first hopping step is 48.
Likewise, the second hopping step enlarges the size of the state space by 48$\times$48.
Sequentially, the size of this configuration space grows exponentially ($48^n$) with the number of hopping steps ($n$).

In order to capture the full effects on local dipoles by a single itinerant carrier, the size of the system needs to be large enough to fully cover the polaronic region.
However, the full diagonalization within the ED approach requires high computational costs, thus limiting the system size one could reach.
%We calculate the system of up to two steps of hopping through the ED approach, which provides the accurate preliminary trend on the growth of the polaron.
%The ED technique is limited in the size of the system due to its high computational cost and the inefficient parallelization of diagonalization.
$N_{\mathbf{p}}$ would saturate with a larger $t/J$ value which corresponds to a larger size of polaron than the system size. 
This effect is clearly shown in Fig.~\ref{fig:dipole_result}(c).
%Therefore, the saturated part of the curves is unreliable.
This ED approach is accurate with a small polaron size in the small $t\ll J$ limit.
%The formation of small polarons and their static properties are used for accuracy tests and comparison to the perturbation approach and the world-line quantum QMC calculations.

We use an in-house C++ code with Linear Algebra PACKage (\textsc{LAPACK}) for the full diagonalization.
The local dipoles are well-ordered along the $\left[111\right]$ direction in the starting configuration.
%At the starting point, all local dipoles are pointed to, which is the lowest energy configuration.
After each hopping step, the configurational energy is calculated by the inter-site dipole-dipole coupling, $E_{\mathrm{config}}=-\sum_{i, i'}\left(J_{ii'}/p_0^{2}\right)\mathbf{p}_i\cdot\mathbf{p}_{i'}$, between dipoles $\mathbf{p}_i=p_0\sum_n \mathbf{\hat{n}} a_{in}^{\dagger} a_{in}$ with site $i$ and $i'$ are all pairs of the nearest neighbors.
An open boundary condition is used in the ED calculation.
Therefore, for the sites at the boundary of the supercell system (3$\times$3$\times$3), the nearest neighbors outside the system are treated to be well-aligned in the symmetry-broken $\left[111\right]$ direction as in FE-ordered state.
%The averaged local dipole moments in the system are then calculated by fully diagonalizing the two hopping steps process, which gives us the wavefunction of the polaron, and therefore other properties such as $N_{\mathbf{p}}$ and the size of the polaron.

% perturbation extrapolation
\begin{figure}[ht]
\centering
\includegraphics[width=0.8\columnwidth]{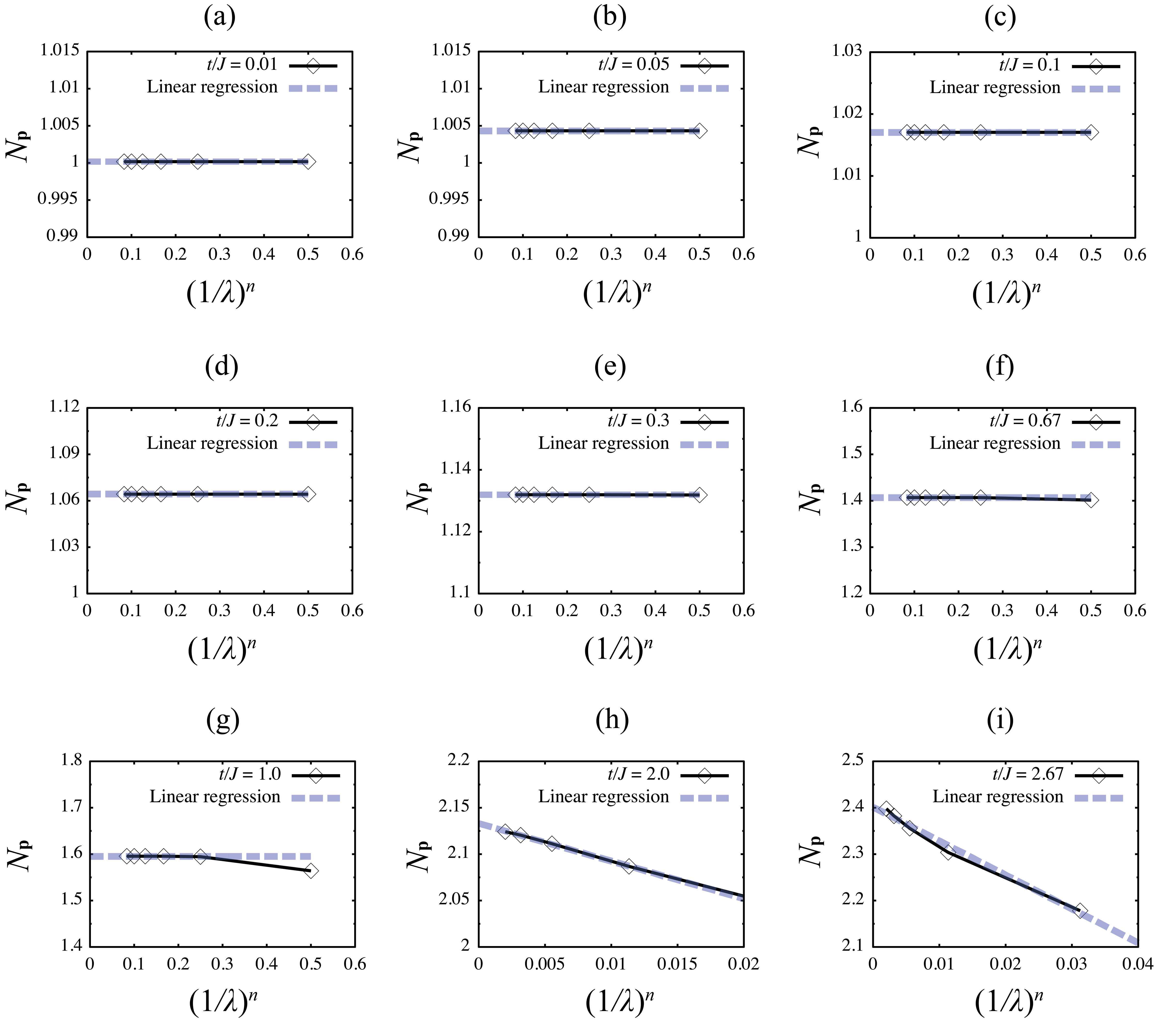}
\caption{
Extrapolation of the average number of suppressed dipoles, $N_{\mathbf{p}}$, against the inverse order, $1/\lambda$, of the perturbation expansion at different $t/J$ values for $\lambda=2,4,6,8,10,12$: (a) $t/J=0.01$, $n=1$; (b) $t/J=0.05$, $n=1$; (c) $t/J=0.1$, $n=1$; (d) $t/J=0.2$, $n=1$; (e) $t/J=0.3$, $n=1$; (f) $t/J=0.67$, $n=1$; (g) $t/J=1.0$, $n=1$; (h) $t/J=2.0$, $n=2.5$; (i) $t/J=2.67$, $n=2.5$.
%; (k) $t/J=3.0$, $n=2.5$; (l) $t/J=4.0$, $n=3.5$
}
\label{fig:extrapolation_diff_t}
\end{figure}

\item \textit{Perturbation theory}: In perturbation calculations, we also start with a clear limit $t\ll J$, in which the hopping term $H_t$ can be treated as a perturbation of the total effective Hamiltonian (Eq.\ref{eqn:total_hamil}).
As the kinetic strength ($t$) is treated as the perturbation term, a larger $t/J$ value naturally requires higher-order term corrections.
The highest order at which the total energy of the polaron converges shows the farthest octahedral site that the carrier would reach in the virtual dynamical process, which also indicates the rough radius ($r_k$) of the polaron.
It is worth noting that one of our assumptions being made, which is the higher energy dynamical process corresponding to the disappearance/emergence of local dipoles, may not be able to be fully absorbed into the Hamiltonian in the $t\gg J$ limit. 
Therefore, other higher-energy physical processes may be of relative importance in this limit.

A convergence test is conducted for higher-order perturbation calculations.
For a ($j+1$)th order perturbation, the energy correction to the total energy is: $\Delta E_{j+1}=E_n^{j+1}=\sum_{m_1}\sum_{m_2}\sum_{m_3}\dots \sum_{m_j} \frac{\left\langle\psi^{(n)}|\hat{V}| \psi^{(m_1)}\right\rangle \left\langle\psi^{(m_1)}|\hat{V}| \psi^{(m_2)}\right\rangle\dots \left\langle\psi^{(m_j)}|\hat{V}| \psi^{(n)}\right\rangle}{\left(E^{(n)}-E^{(m_1)}\right)\left(E^{(n)}-E^{(m_2)}\right)\dots \left(E^{(n)}-E^{(m_j)}\right)}$.
In our study, the average energy change $E^{(m_i)}-E^{(m_{i+1})}$ of each hopping step $i$ is $\sim$4.136$J$ with $t/J$=1.0, and $\left\langle\psi^{(m_i)}|\hat{V}| \psi^{(m_{i+1})}\right\rangle=t$.
Therefore, we can roughly estimate this energy correction at ($j+1$)th order to be proportional to $t^{j+1}/\left((j)!(4.136 J)^{j}\right)$, with the total energy gain with the distance shown in the inset of Fig.~\ref{fig:convergence_size}.
As a natural consequence, at small $t/J\rightarrow 0$, the correction in the energy of each perturbation order ($\Delta E_j$) monotonically decreases with the number of orders ($j$); while $t/J$ is large, the energy correction increases to the maximum value at a certain higher order, then gradually converges with the number of order.
%The above effects are clearly demonstrated in Fig.~\ref{fig:convergence_size}.
%This is naturally the consequence of the coefficient of each perturbation order ($j$), which is proportional to $t^{j}/\left((j-1)!J^{j-1}\right)$.
%Physically, the order at which the total energy converge indicates the rough radius ($r_k$) of the polaron.
As the energy correction at ($j+1$)th order is proportional to $t^{j+1}/\left((j)!(4.136J)^{j}\right)$, the contribution of the infinite order term is always zero as $\lim_{j \to \infty } \frac{t^{j+1}}{\left((j)!(4.136J)^{j}\right)}= 0$.
%This is why we could obtain accurate results of polaron up to $t/J\sim 2.0$.
Note that with the form of $\Delta E_{j}$, $\prod_{m_1}^{m_j}\left(E_n-E_{m_j}\right)=0$ leads to divergence in perturbation calculation.
In our calculation, it is required that the carrier leaves a site with a different local dipole mode from its original mode at each hopping step.
The virtual processes that keep the local dipole in the same direction are reserved in the low-energy subspace of this polaronic Hamiltonian (Eq.\ref{eqn:total_hamil}).

In our calculation, the farthest octahedral site that the virtual dynamic process of the carrier could reach, which is also half of the highest perturbation order, is six formula unit cells ($r_k=6$).
%In the perturbation calculation, the highest perturbation order we could reach is 12th order, which enlarges the state space and allows the formation of polaron up to radius of 6 unit cells length.
(With the perturbation calculation of different orders, we extrapolate the real trend of $N_{\mathbf{p}}$ with $t/J$ as shown in Fig.~\ref{fig:dipole_result}(c).
The linear extrapolation of different $t/J$ values is shown in Fig.~\ref{fig:extrapolation_diff_t}.)
%Generally, the error of the wavefunction normalization in perturbation theory is of the same order of the error due to the lack of higher order terms.
%Therefore, we do not bother to normalize the wavefunction, and that is why the curves seems not suffer from saturation problem due to the finite system size as shown in Fig.~\ref{fig:dipole_result}.
%According to the $N_{\mathbf{p}}$ curves in Fig.~\ref{fig:dipole_result}(b), a clear phase transition point occurs at a critical $(t/J)_c$ value, after which the size of the polaron exponentially grows to the whole system size.
%This could explain the FE-PE phase transitions occur at a relative low doping density ($\delta_\textrm{c2}\sim 10\%$) in a wide range of experiments~\cite{kolodiazhnyi_PRL_2010,fujioka_2015,cordero_metal_2019, HXu_PRB2016,Benedek_JMC2016,Lei_nanolett2018,Sergienko_prl2004}.
We use an in-house C++ code for the perturbation calculations, with a 7$\times$7$\times$7 supercell size.
The size of the supercell is large enough as the highest order we reach is the 12th order, which indicates 6 outward hopping steps at maximum.
A periodic boundary of the system is used for the virtual hopping process.
The energy of each configuration is exactly the same as in ED calculations, calculated as $E_{\mathrm{config}}=-\sum_{i,i'}\left(J_{ii'}/p_0^{2}\right)\mathbf{p}_i\cdot\mathbf{p}_{i'}$.
It is worth noting that, the carrier can hop to every neighboring site at each step as long as it returns to the original doped site after 12 hopping steps.
Therefore, the carrier's trajectory could form loops, or revisit a single site multiple times without restriction.

\item \textit{World-line quantum Monte Carlo (QMC)}: We use QMC to measure the thermal average of the FE order $\langle \mathrm{\mathbf{O}}\rangle$ at finite temperatures. 
The $H_{R}$ term in Hamiltonian (Eq.\ref{eqn:total_hamil}) is not included in order to study the quantum effect only. 
The FE order can be written as $\langle \mathrm{\mathbf{O}}\rangle=\frac 1 Z \Tr (\mathrm{\mathbf{O}}\ \mathrm{exp} (-\beta H) )$, where $H$ is the Hamiltonian, $\beta=1/(k_B T)$ is the inverse temperature, and $Z=\Tr (\mathrm{exp}(-\beta H))$ is the partition function.

In our QMC calculation, the basis is taken as $c_i^\dagger\prod_{j}a^\dagger_{j n_j}|0\rangle$. The dimension of Hilbert space is $8^{L^3-1}\times L$ where $L$ is the system size. 
The Hamiltonian can be written in terms of $H=T+H_0$, where $H_0$ is the diagonal part and $T$ is the non-diagonal part. 
Using an interaction picture, we can represent the Boltzmann factor as $\exp{(-\beta H)}=\exp{(-\beta H_0)}\sum_{k=0}^\infty (-1)^k \int_0^\beta \,\mathrm{d}\tau_1\int_0^{\tau_1}\,\mathrm{d}\tau_2\cdots\int_0^{\tau_{k-1}}\,\mathrm{d}\tau_k V_{\tau_1}^D V_{\tau_2}^D\cdots V_{\tau_k}^D$ where $V_{\tau_i}^D=\exp{(\tau_i H_0)}V\exp{(-\tau_i H_0)}$. 
Here we treat $\tau$ as the imaginary time and plug complete relations between each $V_{\tau_i}^D$, then $\langle \mathrm{\mathbf{O}}\rangle$ can be treated as a weighted average among all possible closed paths (world line) in space-time. The non-diagonal term (the kinetic part) leads to a ``hop'' between lattice sites, thus creating a ``kink'' on the world line. 
Importance sampling is done through the Metropolis algorithm. Ergodicity and detailed balance ensure the sequence of world lines converges to the desired distribution. 
%In principle, a global update scheme must be applied to generate different topological structures of paths. 
%However, in our study, the additional freedom in the dipole orientation makes it hard to generate a non-trivial topological structure in the phase space, because the electron must recover the dipole configuration after it comes back. Therefore, local update schemes are already a very good approximation, which consists of two schemes-- creation/annihilation of ``kink-antikink" pair and the time shift of kinks. 

We use an in-house MATLAB code for the world-line QMC calculations, with a $5\times5\times5$ supercell under periodic boundary conditions. 
The supercell size used was tested by a system size scaling as shown in Fig.~\ref{fig:supple_QMC}(a).
%Same to the above two approaches, all local dipoles are aligned with a single carrier introduced as the lowest energy starting configuration. 
At every kink point, we store the imaginary time, the electron position, and the dipole configuration as matrices, and build a mapping between them. The weight of the paths and the estimator can be calculated from these matrices. 
In the update scheme, we generate a random kink-antikink pair at $\tau_1$ and $\tau_2$. Between these two time points, we propose a carrier hopping to one of the neighboring sites, randomly leaving a dipole orientation at the original site. Then, we shift the kink from $\tau_1$ to $\tau_2$ correspondingly. 
We accept these updates with certain probabilities obtained from the detailed balance. 
Through this scheme, the system gradually converges to its thermally equilibrated state.
The calculation only converges fast at high temperatures $(k_B T)/J>0.1$, below which the acceptance ratio becomes exponentially small. 
%In the region of $(k_B T)<J$, the broken inversion symmetry is well preserved in a long timescale, because it consumes considerable energy to globally flip the local dipoles. 

%Different from the ground state study of the formation of the polaron, the finite temperature calculations provide us with a thermodynamically equilibrated Bloch-state, which involves the effect of the low-energy kinetic process of the polaron as a quasi-particle. 
%The further effect of the polaron ``tail" to the FE order and the lack of the thermal effects to self-correct the local dipoles lead to a larger $N_{\mathbf{p}}$ in QMC simulations compared to those obtained from ED and perturbation approaches.
As shown in Fig.~\ref{fig:supple_QMC}(b), $N_{\mathbf{p}}$ slightly decreases as $T$ increases in the low-$T$ range.
%This slight Np decrease is because with the increase of $T$, the thermal distribution suppresses the effect of the virtual kinetic process by averaging thermodynamically equilibrated Bloch states into a localized-like state, therefore $N_{\mathbf{p}}$ slightly decreases with temperature.
The slight decrease in $N_{\mathbf{p}}$ is attributed to the rise in temperature, which causes the thermal distribution to suppress the impact of the virtual kinetic process by averaging out thermodynamically equilibrated Bloch states into a localized state.
Fig.~\ref{fig:supple_QMC}(b) also showcases the dominance of thermal fluctuations over quantum fluctuations in the high-temperature range.
%demonstrates the dominance of the thermal fluctuation over the quantum fluctuation in the relative high-$T$ range.
%Already defined Np
%The projection of dipoles along [111] direction is statistically averaged in our simulation, as a measurement of polaron size at finite temperature.

\begin{figure}[!h]
\includegraphics[width=0.8\columnwidth]{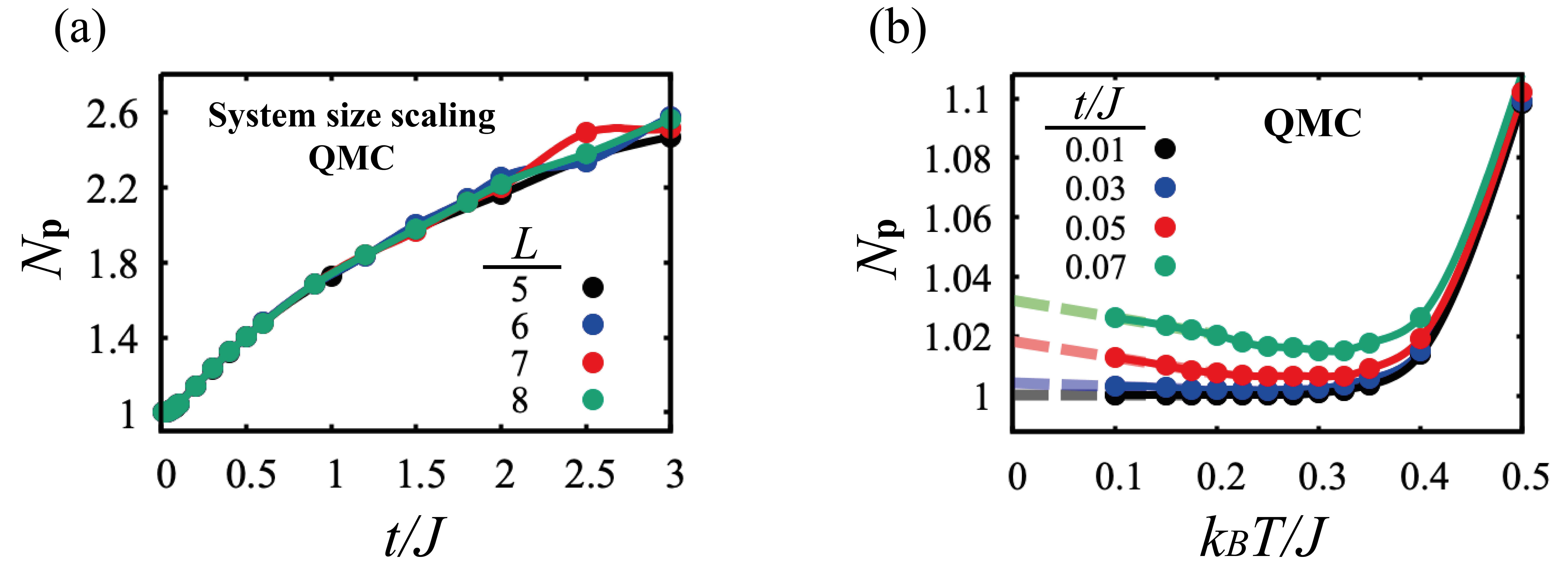}
\centering
\caption{
(a) The $t/J$ dependence of $N_{\mathbf{p}}$ at various cubic system size $L^3$ for $L=5,6,7,8$ under a fixed finite temperature $k_{B}T/J=0.2$; (b) Temperature dependence of $N_{\mathbf{p}}$ for different $t/J$ parameter.
}
\label{fig:supple_QMC}
\end{figure}

\item \textit{Estimating the renormalized kinetic strength}:

When the ground state of the polaron is energetically well separated from the excited states, the renormalized kinetic energy of the polaron is basically $\widetilde{t}_{jj'}=\langle\psi_{j}|H| \psi_{j'}\rangle \approx \langle\psi_{j}|H_t| \psi_{j'}\rangle$, where $\mid \psi_{j}\rangle$ and $\mid \psi_{j'}\rangle$ denote the polaronic ground states centered at nearby sites, $j$ and $j'$, respectively.
(For larger polarons, correction due to lack of orthogonality between $\mid \psi_{j}\rangle$'s might introduce additional correction.)
Note that $\mid \psi_{j}\rangle$ is a many-body states including not only the carrier, but also all the electric dipoles in the system.
It is therefore easy to see why the polaron generically becomes very heavy: unless the electron first dynamically visits all the dipoles in the back side of the polaron and happens to leave them along the FE-ordered direction, the polaron cannot move its center forward.
As an example, in the small kinetic region, say $t/J=0.1$, we found the renormalized $\widetilde{t}_{jj'}=0.012t$ is easily suppressed by more than two orders of magnitude. 

\end{enumerate}
%\nolinenumbers

\bibliographystylemethod{plain}
\bibliographymethod{polaron_paper}

\section*{Acknowledgements}

%Acknowledgements should be brief, and should not include thanks to anonymous referees and editors, or effusive comments. Grant or contribution numbers may be acknowledged.
We acknowledge helpful discussions with Wei Wang, Chi-Ming Yim, and Anthony Charles Hegg.
This work is supported by the National Natural Science Foundation of China (NSFC)	\#11674220 and \#12042507.
We also acknowledge the support from the International Postdoctoral Exchange Fellowship Program (YJ20210137) by the Office of China Postdoc Council (OCPC).

\section*{Author contributions statement}

W.K. and F.G. conceived the idea of the project. 
F.G., J.W. and Z.L. developed the model through exact diagonalization, perturbation and quantum Monte Carlo  approaches.
F.G. performed the ED calculations. F.G. and Z.L. performed the perturbation calculations.
J.W. performed the quantum Monte Carlo calculations.
F.G. and W.K. analyzed the results and wrote the paper.
All authors discussed the results and contributed to revising and editing the manuscript.
W.K. supervised the project.

\section*{Additional information}

To include, in this order: \textbf{Accession codes} (where applicable); \textbf{Competing interests} (mandatory statement). 

The corresponding author is responsible for submitting a \href{http://www.nature.com/srep/policies/index.html#competing}{competing interests statement} on behalf of all authors of the paper. This statement must be included in the submitted article file.

\end{document}